\documentclass{PoS}
\usepackage{graphicx}
\usepackage{amsmath}
\usepackage{amsfonts}
\usepackage{amssymb}
\usepackage{epsfig}
\usepackage{times}
\usepackage{cite}
\usepackage{calc}
\usepackage{version}
\usepackage[french,english]{babel}

\title{Mean multiplicity of light and heavy quark initiated jets}

\ShortTitle{Hadronic multiplicities}

\author{\speaker{Redamy P\'erez-Ramos}\\\\
        Departament de Fisica Te\`orica and IFIC, Universitat de Val\`encia-CSIC\\
        Dr. Moliner 50, E-46100 Burjassot, Spain\\
        E-mail: \email{redamy.perez@uv.es}}

\abstract{After inserting the heavy quark mass 
dependence into QCD partonic evolution
equations, we determine the mean charged hadron
multiplicity of jets produced in high 
energy collisions. We thereby extend the so-called dead cone effect to the 
phenomenology of multiparticle production in QCD jets and find that the
average multiplicity of heavy-quark initiated 
jets decreases significantly as compared to the massless case, 
even taking into account
the weak decay products of the leading primary quark. 
We emphasize the relevance of our study 
as a complementary check of $b$-tagging techniques
at hadron colliders like the Tevatron and the LHC.}

\FullConference{Light Cone 2010 - LC2010\\
		June 14-18, 2010\\
		Valencia, Spain}

\begin{document}

\section{Introduction}

High-$p_t$ jets can be initiated either in a short-distance
interaction among partons in high energy collisions such as $pp$, 
$p\bar p$, in the DIS $e^\pm p$, the $e^+e^-$ annihilation and via
electroweak (or new physics) processes. 
One well-known example
is given by the decay chain of the top quark 
$t\ \to\ H^+\ b$,  
where the $b$ quark should start a jet. Thus the ability to
identify jets from the fragmentation and hadronization
of $b$ quarks becomes very important for such Higgs boson searches.
Needless to say, the relevance of $b$-tagging extends over 
many other channels in the quest for new physics at 
hadron colliders. The experimental 
identification of $b$-jets relies upon several of their properties
in order to reject background, e.g. 
jets initiated by lighter quarks or gluons. First,
the fragmentation is hard and the leading $b$-hadron retains
a large part of the original $b$ quark momentum. In addition, 
the weak decay products may have a large transverse momentum with respect to
the jet axis therefore allowing separation from the rest of 
the cascade particles. 
Lastly, the relatively long lifetime
of $b$-hadrons leading to displaced vertices which can be
identified by using well-known impact parameter techniques \cite{Aad:2009wy}.   
Still, a fraction of light jets could be
mis-identified as $b$-jets, especially at large 
transverse momentum of the jet. Now, let us point out that an essential difference 
between heavy and light quark jets results from
kinematics constraints: the gluon radiation
off a quark of mass $m$ and energy $E\gg m$ is suppressed
inside a forward cone with an opening angle $\Theta_m=m/E$,
the so-called {\em dead cone} phenomenon \cite{Dokshitzer:1991fd,Dokshitzer:2005ri}.

In this work, we compute the average (charged) multiplicity 
of a jet initiated by a heavy quark. For this purpose,
we extend the modified leading logarithmic approximation (MLLA) 
evolution equations \cite{Dokshitzer:1991wu} to the case where the 
jet is initiated by a heavy (charm, bottom) quark. 
The average multiplicity of light quarks jets produced in high energy collisions can be 
written as $N(Y)\propto\exp\left\{\int^Y\gamma(y)dy\right\}$ ($Y$ and $y$ are
defined in section \ref{sec:kivar}), where $\gamma\simeq\gamma_0+\Delta\gamma$ 
is the anomalous dimension that accounts for
soft and collinear gluons in the double logarithmic 
approximation (DLA) $\gamma_0\simeq\sqrt{\alpha_s}$, in addition to
hard collinear gluons $\Delta\gamma\simeq\alpha_s$ or single logarithms (SLs),
which better account for energy conservation and the running of the 
coupling constant $\alpha_s$ \cite{Dokshitzer:1991wu}. 

In the present work we evaluate the mean multiplicity
as a function of the mass of the heavy quark. 
Actually, the mass depends on the scale $Q$ of the hard process 
in which the heavy quark participates. 
In the $\overline{MS}$ renormalization scheme, for example,  
the running mass becomes a function of the
strong coupling constant $\alpha_s(Q)$,
and the pole mass defined through 
the renormalized
heavy quark propagator (for a review see \cite{Kluth:2006bw} and references therein). 
In order to assess the scheme-dependence of
our calculations beyond leading-order
we have considered a broad range of possible values for 
the charm and bottom masses:   
$m_c=1-1.5$ GeV and $m_b=3-5$ GeV, respectively \cite{Kluth:2006bw}.
Lastly, we will see that under the assumption of local parton 
hadron duality (LPHD)
as hadronization model \cite{Azimov:1984np,Dokshitzer:1995ev}, 
light- and heavy-quark initiated jets show significant differences
regarding particle multiplicities as a consequence of soft gluon suppression inside
the dead cone. Such differences could be exploited
by using auxiliary criteria complementing $b$-tagging procedures to
be applied to jets with very large transverse momentum, as advocated
in this study.

\newpage

\section{Kinematics and variables}
\label{sec:kivar}
As known from jet calculus for light quarks, 
the evolution time parameter determining the structure
of the parton branching of the primary gluon is given by 
(for a review see \cite{Dokshitzer:1991wu} and references therein)
\begin{equation}\label{eq:masslessev}
y=\ln\left(\frac{k_\perp}{Q_0}\right),\quad k_\perp=zQ\geq Q_0,
\quad Q=E\Theta\geq Q_0,
\end{equation}
where $k_\perp$ is the transverse momentum of the gluon emitted off the
light quark, $Q$ is the virtuality of the jet (or jet hardness), 
$E$ the energy of the 
leading parton, $Q_0/E\leq\Theta\leq\Theta_0$  
is the emission angle of the gluon ($\Theta\ll1$), $\Theta_0$ the 
total half opening angle of the jet being 
fixed by experimental requirements, 
and $Q_0$ is the collinear cut-off parameter.
Let us define in this context the variable $Y$ as $y=Y+\ln z,\; Y=\ln\left(\frac{Q}{Q_0}\right)$.
The appearance of this scale is a consequence of angular ordering
(AO) of successive
parton branchings in QCD cascades \cite{Azimov:1984np,Dokshitzer:1991wu}. 
An important difference in the structure of light ($\ell\equiv q=u,d,s$) 
versus heavy quark ($h\equiv Q=c,b$) jets stems from the dynamical 
restriction on the phase space of primary gluon radiation in the heavy quark 
case, where the gluon radiation off an energetic quark $Q$ with mass
$m$ and energy $E\gg m$ is suppressed inside the forward cone with an
opening angle $\Theta_m=m/E$, the above-mentioned
dead cone phenomenon \cite{Dokshitzer:2005ri}. 

The corresponding evolution time parameter for a jet initiated 
by a heavy quark with energy $E$ and mass $m$ appears in a natural 
way and reads $\tilde y=\ln\left(\frac{\kappa_\perp}{Q_0}\right),\;
\kappa_\perp^2=k_\perp^2+z^2m^2$ \cite{Dokshitzer:2005ri},
which for collinear emissions $\Theta\ll1$ can also be rewritten in the form
\begin{equation}\label{eq:Qtheta}
\kappa_\perp=z\tilde Q,\quad \tilde Q=E\left(\Theta^2+\Theta_m^2\right)^{\frac12},
\end{equation}
with $\Theta\geq\Theta_m$ (see Fig.\,\ref{fig:Qsplit}). 
An additional comment is in order concerning the 
AO for gluons emitted off the heavy quark. In (\ref{eq:Qtheta}), $\Theta$ is the  
emission angle of the primary gluon $g$ being emitted off the heavy quark. Now
let $\Theta'$ be the emission angle of a second gluon $g'$ relative to the primary 
gluon with energy $\omega'\ll\omega$ and $\Theta''$ the emission angle relative to 
the heavy quark; in this case 
the {\em incoherence} condition $\Theta'^2\leq(\Theta^2+\Theta_m^2)$ together with 
$\Theta''>\Theta_m$ (the emission angle of the second gluon should still be larger 
than the dead cone) naturally leads (\ref{eq:Qtheta}) 
to become the proper evolution
parameter for the gluon subjet (for more details see \cite{Dokshitzer:2005ri}). 
For $\Theta_m=0$, the standard 
AO ($\Theta'\leq\Theta$) is recovered.
Therefore, for a massless quark, the virtuality of the
jet simply reduces to $Q=E\Theta$ as given above.
The same quantity $\kappa_\perp$ determines the scale of the running coupling
$\alpha_s$ in the gluon emission off the heavy quark. It can be related to the anomalous
dimension of the process by
\begin{equation}
\gamma_0^2(\kappa_\perp)=
2N_c\frac{\alpha_s(\kappa_\perp)}{\pi}=\frac1{\beta_0(\tilde y+\lambda)},\quad
\beta_0(n_f)=\frac1{4N_c}\left(\frac{11}3N_c-\frac23n_f\right),\quad 
\lambda=\ln\frac{Q_0}{\Lambda_{QCD}},
\end{equation}
where $n_f$ is the number of active flavours and $N_c$ the number of colours.
The variation of the effective 
coupling $\alpha_s$ as $n_f\to n_f+1$ over the heavy quarks threshold has 
been suggested by next-to-leading (NLO) calculations in the $\overline{MS}$ scheme 
\cite{Dokshitzer:1995ev} and is sub-leading in this frame. 
In this context $\beta_0(n_f)$ will be evaluated at the total number of quarks we consider in our application. The four scales of the process are related as follows,
$$
\tilde Q\gg m\gg Q_0\sim\Lambda_{QCD},
$$
where $Q_0\sim\Lambda_{QCD}$ corresponds to the limiting spectrum 
approximation \cite{Dokshitzer:1991wu}. Finally, the dead cone 
phenomenon imposes the following bounds of integration to the perturbative regime
\begin{equation}\label{eq:bounds}
\frac{m}{\tilde Q}\leq z\leq 1-\frac{m}{\tilde Q},\quad
m^2\leq\tilde Q^2\leq E^2(\Theta_0^2+\Theta_m^2),
\end{equation} 
which now account for the phase-space of the heavy quark jet. The last inequality
states that the minimal transverse momentum of the jet $\tilde Q=E\Theta_m=m$ 
is given by the mass of the heavy quark, which enters the game as the natural 
cut-off parameter of the perturbative approach.
\section{Definitions and notation}
\label{sec:defandnot}
The multiplicity distribution is defined by the formula
\begin{equation}
P_n=\frac{\sigma_n}{\sum_{n=0}^{\infty}\sigma_n}
=\frac{\sigma_n}{\sigma_{inel}},\quad \sum_{n=0}^{\infty}P_n=1
\end{equation}
where $\sigma_n$ denotes the cross section of an $n$-particle yield process, 
$\sigma_{inel}$ is the inelastic cross-section, and the sum runs 
over all possible values of $n$.

It is often more convenient to represent multiplicity distributions by 
their moments. All such sets can be 
obtained from the generating functional
$Z(y,u)$ \cite{Dokshitzer:1991wu} defined by 
$$
Z(y,u)=\sum_{n=0}^{\infty}P_n(y)\ (1+u)^n
$$
at the energy scale $y$. For fixed $y$, we can drop this variable 
from the {\em azimuthally} 
averaged generating functional $Z(u)$. The average multiplicity 
is defined by the formula,
\begin{equation}\label{eq:meann}
\left<n\right>\equiv N=\sum_{n=0}^{\infty}P_nn,\quad
P_n=\frac{1}{n!}\frac{d^nZ(u)}{du^n}\Big|_{u=-1}. 
\end{equation}
We will compute the average multiplicity (\ref{eq:meann})
of partons in jets to be denoted hereafter as $N_A$, with $A=Q,q,g$, 
corresponding to a heavy, light quark or gluon initiated jet
respectively. 

Once arrived at his point, let us make an important distinction
between two different particle sources populating
heavy-quark initiated jets. 
On the one hand, parton cascade from gluon emission
yields the QCD component of the total jet multiplicity
(the main object of our present study), {\em excluding weak
decay products of the leading primary quark}
at the final stage of hadronization. On the other hand,
the latter products coming from the leading flavoured hadron
should be taken into account in the measured multiplicities
of jets. We shall denote the average charged hadron multiplicity
from the latter source as $N_A^{dc}$. Hence
the total charged average multiplicity, $N_A^{total}$, reads
\begin{equation}\label{eq:total}
N_A^{total}=N^{ch}_A+N_A^{dc}\ ;\quad A=q,Q.
\end{equation} 
As a consequence of the LPHD, $N^{ch}_A={\cal K}^{ch}\times N_A$ 
\cite{Dokshitzer:1995ev,Azimov:1984np}, where
the free parameter ${\cal K}^{ch}$ normalizes the average multiplicity
of partons to the average multiplicity of charged hadrons.
For charm and bottom quarks, we will respectively set the values
$N_c^{dc}=2.60 \pm 0.15$ and $N_b^{dc}=5.55 \pm 0.09$
\cite{Dokshitzer:2005ri,Akers:1995ww}, while in light quark jets one expects
$N_q^{dc}=1.2 \pm 0.1$ \cite{:1994qa}. In this work, we advocate
the use of such a difference between average jet multiplicities as   
a signature to distinguish {\em a posteriori}
heavy from light quark jets, particularly in $b$-tagging techniques
applied to the analysis of many interesting decay channels. 

\section{QCD evolution equations}
The splitting functions 
\begin{equation}
P(z,\alpha_s)=\alpha_sP^{(0)}(z)+\alpha_s^2P^{(1)}(z)+\ldots
\end{equation}
where $P^{(0)}(z)$ and $P^{(1)}(z)$ are respectively the LO and NLO 
splitting functions, can be associated to each vertex of 
the process in the partonic shower. $P(z,\alpha_s)$
determines the decay probability of a parent parton (quark, anti-quark, gluon) into
two offspring partons of energy fractions $z$ and $1-z$. In this work, we are rather 
concerned with calculations which only involve the LO $P^{(0)}(z)$ 
splitting functions in the evolution equations \cite{Dokshitzer:1991wu}. 
\begin{figure}[h]
\begin{center}
\epsfig{file=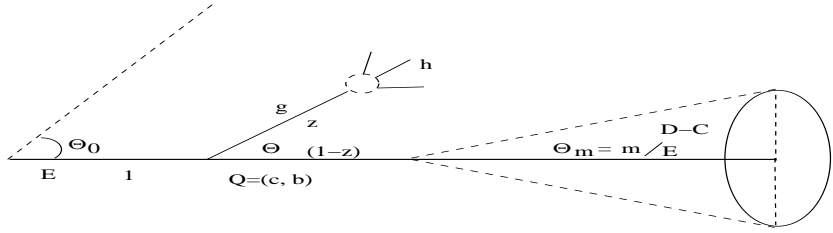, height=3truecm,width=11truecm}
\caption{\label{fig:Qsplit} Parton splitting in the process $Q\to Qg$: 
a {\em dead cone} with opening angle $\Theta_m$
is schematically shown.}  
\end{center}
\end{figure}

Let us start by considering the
the splitting process, $Q\to Qg$, $Q$ being a heavy quark and $g$ 
the emitted gluon which is displayed in Fig.\ref{fig:Qsplit}; the corresponding LO
splitting function reads \cite{Baier:1973ms,Krauss:2003cr}
\begin{equation}\label{eq:PQg}
P_{Qg}^{(0)}(z)=
\frac{C_F}{N_c}\left[\frac{1}{z}-1+\frac{z}2-
\frac{z(1-z)m^2}{k_\perp^2 + z^2m^2}\right],\quad
P_{QQ}^{(0)}(z)=P_{Qg}^{(0)}(1-z)
\end{equation}
where $k_\perp\approx min(zE\Theta,(1-z)E\Theta)$ is the transverse 
momentum of the soft gluon being
emitted off the heavy quark. The previous formula (\ref{eq:PQg}) has the 
following physical interpretation, for $k_\perp\ll z^2m^2$, the corresponding
limit reads $P_{Qg}(z)\to\frac{C_F}{2N_c}z$ and that is why, at leading logarithmic
approximation, the forward emission of soft and collinear gluons off the heavy 
quark becomes suppressed once $\Theta\ll\Theta_m$, while the
emission of hard and collinear gluons dominates in this region.

For the massless process $g\to gg$, we adopt the standard three gluon 
vertex kernel \cite{Dokshitzer:1991wu,Dremin:2000ep}
\begin{equation}\label{eq:Pgg}
P^{(0)}_{gg}(z)=\frac{1}{z}-(1-z)[2-z(1-z)],
\end{equation}
and finally for $g\to Q\bar Q$, we take \cite{Baier:1973ms,Krauss:2003cr}
\begin{equation}\label{eq:PgQ}
P_{gQ}^{(0)}(z)=\frac1{4N_c}\left[1-2z(1-z)+\frac{2z(1-z)m^2}{k_\perp^2+m^2}\right],
\end{equation}
which needs to be resummed together with the three gluon vertex contribution. 
However, as a first approach to this problem, we neglect the production
of heavy quark pairs inside gluon and quark jets, making use of 
\cite{Dokshitzer:1991wu,Dremin:2000ep}
\begin{equation}
P^{(0)}_{gq}(z)=P^{(0)}_{gQ}(z)|_{m=0}=\frac1{4N_c}\left[1-2z(1-z)\right].
\end{equation}
Including mass effects in the evolution equations also requires the replacement
of the massless quark propagator $1/k_\perp^2$ by the massive quark propagator
$1/(k_\perp^2+z^2m^2)$ \cite{Krauss:2003cr}, 
such that the phase space for soft and collinear 
gluon emissions off the heavy quark can be written 
in the form \cite{Krauss:2003cr}
\begin{equation}\label{eq:jetcs}
d^2\sigma_{Qg}\simeq
\gamma_0^2
\frac{dk_\perp^2}{k_\perp^2+z^2m^2}
dzP_{Qg}^{(0)}(z),
\end{equation}
where $P_{Qg}^{(0)}(z)$ is given by (\ref{eq:PQg}). Working out the structure
of (\ref{eq:PQg}) and setting $k_\perp\approx zE\Theta$ one has
\begin{equation}\label{eq:PQgbis}
P_{Qg}^{(0)}(z)=
\frac{C_F}{N_c}\left[\frac{1}{z}\frac{\Theta^2}{\Theta^2+\Theta_m^2}-1+\frac{z}2
+\frac{\Theta_m^2}{\Theta^2+\Theta_m^2}\right].
\end{equation}
The system of QCD evolution 
equations for the heavy quark initiated jet is found to read \cite{Ramos:2010cm}
\begin{eqnarray}
\frac{N_c}{C_F}\frac{dN_Q}{d\tilde Y}\!\!&\!\!=\!\!&\!\!
\epsilon_1(\tilde Y,L_m)\int_{\tilde Y_{m}}^{\tilde Y_{ev}}
d\tilde y\gamma_0^2(\tilde y)N_g(\tilde y)-\epsilon_2(\tilde Y,L_m)\gamma_0^2(\tilde Y)N_g(\tilde Y),
\label{eq:NQhter}\\
\frac{dN_g}{d\tilde Y}\!\!&\!\!=\!\!&\!\!
\int_{\tilde Y_{m}}^{\tilde Y_{ev}}d\tilde y\gamma_0^2(\tilde y)N_g(\tilde y)-A(\tilde Y,L_m)
\gamma_0^2(\tilde Y)N_g(\tilde Y),
\label{eq:NGhter}
\end{eqnarray}
where
\begin{equation}\label{eq:Anf}
A(\tilde Y,L_m)=a(n_f)-\left[2+\frac{n_f}{2N_c}
\left(1-2\frac{C_F}{N_c}\right)\right]e^{-\tilde Y+L_m}+
\frac12\left[1+\frac{n_f}{N_c}\left(1-2\frac{C_F}{N_c}\right)\right]e^{-2\tilde Y+2L_m},
\end{equation}
with
\begin{equation}
a(n_f)=\frac{1}{4N_c}\left[\frac{11}3N_c+\frac23n_f
\left(1-2\frac{C_F}{N_c}\right)\right],\quad \tilde Y_{m}\equiv L_m=\ln\frac{m}{Q_0},
\end{equation}
and
\begin{equation}\label{eq:epsilon12}
\epsilon_1(\tilde Y,L_m)=1-e^{-2\tilde Y+2L_m},\quad 
\epsilon_2(\tilde Y,L_m)=\frac34-\frac{3}{2}
e^{-\tilde Y+L_m}-e^{-2\tilde Y+2L_m}.
\end{equation}
There are the following kinds of power suppressed corrections to the heavy quark multiplicity: 
the leading integral term of
(\ref{eq:NQhter}) is ${\cal O}(\frac{m^2}{\tilde Q^2})$ suppressed, while subleading MLLA 
corrections appear in the standard form ${\cal O}(\sqrt{\alpha_s})$ like in the massless case, finally
${\cal O}(\frac{m}{\tilde Q}\sqrt{\alpha_s})$ and
${\cal O}(\frac{m^2}{\tilde Q^2}\sqrt{\alpha_s})$, which are new in this context.
Massless results can be recovered after setting $m/\tilde Q\to Q_0/Q$ in 
(\ref{eq:Anf}) and (\ref{eq:epsilon12}). 
For massive particles however, these terms are somewhat larger
and can not be neglected in our approach unless they are evaluated for much 
higher energies than at present colliders. 
On top of that, the corresponding massless equations in the high energy limit are obtained 
from (\ref{eq:NQhter}) and (\ref{eq:NGhter}) simply by setting $\tilde y\to y$, $\tilde Y\to Y$, 
$Y_{ev}\to Y$, $Y_{m}\to 0$, $\epsilon_1\to1$,
$$
\epsilon_2\to\tilde\epsilon_2=\frac34-\frac32e^{-Y}+{\cal O}\left(e^{-2Y}\right),
\quad A\to\tilde A=a(n_f)-\left[2+\frac{n_f}{2N_c}
\left(1-2\frac{C_F}{N_c}\right)\right]e^{-Y}+{\cal O}\left(e^{-2Y}\right),
$$
and are written in the standard form \cite{Dremin:2000ep}
\begin{eqnarray}\label{eq:Nqlight}
\frac{N_c}{C_F}\frac{dN_q}{dY}=
\int_{0}^{Y}
d y\gamma_0^2N_g(y)-\tilde\epsilon_2\gamma_0^2N_g(Y),\;
\frac{dN_g}{dY}=
\int_{0}^{Y}dy\gamma_0^2N_g(y)-\tilde A
\gamma_0^2N_g(Y),
\end{eqnarray} 
with the initial condition $N_{g,q}(Y=0)=1$ at threshold.
Notice that (\ref{eq:NQhter}) and (\ref{eq:NGhter}) are valid only for $m\gg Q_0$ and
therefore $m\to 0$ does not reproduce the correct limit, which has to be smooth as given 
by the massless equations (\ref{eq:Nqlight}).
Since heavy quarks are less sensitive to recoil effects, 
the subtraction terms $\propto e^{-\tilde Y+L_m}$ and $\propto e^{-2\tilde Y+2L_m}$ 
in $\epsilon_2(\tilde \tilde Y,L_m)$ diminish
the role of energy conservation as compared to massless quark initiated jets.
As a consistency check, upon integration over $\tilde Y$ of
the DLA term in Eq.(\ref{eq:NQhter}), the phase space structure of 
the radiated quanta reads
\begin{equation}\label{eq:dconesimple}
N_Q(\ln\tilde Q)\approx1+\frac{C_F}{N_c}\int_{0}^{\Theta^2_0}
\frac{\Theta^2d\Theta^2}{(\Theta^2+\Theta_m^2)^2}\int_{m/\tilde Q}^{1-{m/\tilde Q}}\frac{dz}{z}
\left[\gamma_0^2N_g\right](\ln z\tilde Q).
\end{equation}
Notice that the lower bound over $\Theta^2$ in (\ref{eq:dconesimple}) ($\tilde Y$ in (\ref{eq:NQhter}))
can be taken down to ``0" ($Y_m=L_m$ in (\ref{eq:NQhter})) because the heavy quark mass plays the 
role of collinear cut-off parameter.
\section{Phenomenological consequences}

\begin{figure}[h]
\begin{center}
\epsfig{file=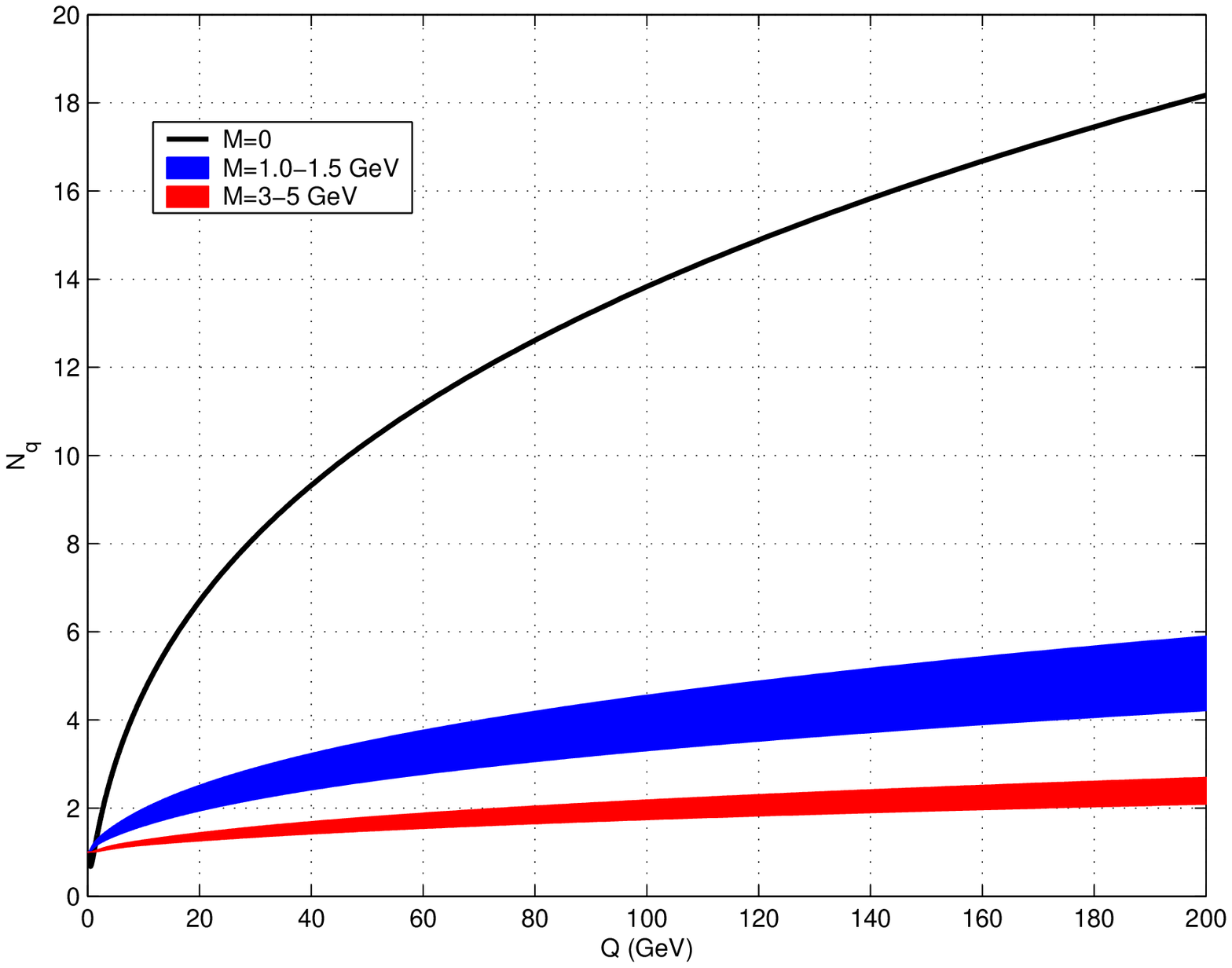, height=5truecm,width=6.5truecm}
\hskip 0.5cm
\epsfig{file=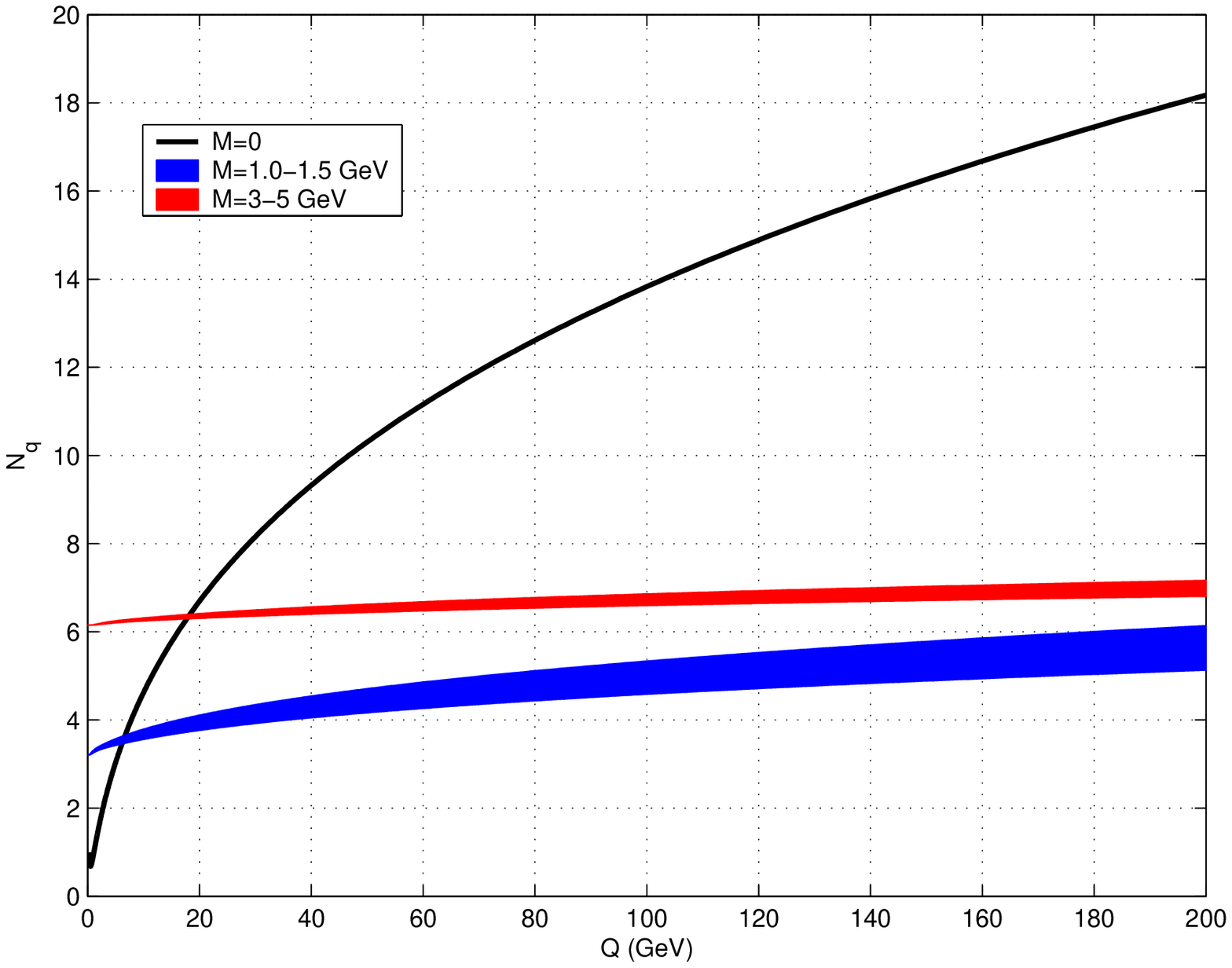, height=5truecm,width=6.5truecm}
\caption{\label{fig:NQ} Massless 
and massive quark jet average multiplicity $N_Q$ for the primary state radiation (left)
and $N_Q^{total}$ including heavy quark flavour decays (right)
as a function of the jet hardness $Q$. Bands indicate $m_c$ and $m_b$ in the $[1,1.5]$ and 
$[3,5]$ GeV intervals respectively.}  
\end{center}
\end{figure}

Working out the structure of (\ref{eq:NQhter}) and (\ref{eq:NGhter}), we obtain
the rough difference between the light and heavy quark jet multiplicities, which yields,
\begin{equation}\label{eq:Nqdiff}
N_q-N_Q\stackrel{E\to\infty}{\approx}\left[1-\exp\left(-2\sqrt{\frac{L_m}{\beta_0}}\right)\right]N_q,\quad
N_q\propto\exp2\sqrt{\frac{\tilde Y}{\beta_0}}.
\end{equation}
It can be seen that (\ref{eq:Nqdiff}) is exponentially increasing because it is dominated by
the leading DLA energy dependence of $N_q$. According to (\ref{eq:Nqdiff}), the gap 
arising from the dead cone effect should be bigger for the $b$ than for the $c$ quark
at the primary state bremsstrahlung radiation off the heavy quark jet.
The approximated solution of the 
evolution equations leads to the rough behaviour of $N_q-N_Q$ in 
(\ref{eq:Nqdiff}), which is not exact in its present form. In Fig.\ref{fig:NQ} (left), we display
the numerical solution of the evolution equations (\ref{eq:Nqlight}) for $N_q$ and
(\ref{eq:NQhter}) for $N_Q$ corresponding to the heavy quark mass intervals 
$m_c=1-1.5$ GeV, $m_b=3-5$ GeV.
Let us remark that the gap arising between the light quark jet multiplicity and
the heavy quark jet multiplicity follows the trends given
by (\ref{eq:Nqdiff}) asymptotically with $E\to\infty$. Finally, 
as expected for massless quarks $L_m=0$, the  difference $N_q-N_Q$ vanishes.
In this study we advocate the role of 
mean multiplicities of jets as a
potentially useful signature for $b$-tagging and 
new associated physics \cite{SanchisLozano:2008te} 
when combined with other selection criteria.
In Fig.\,\ref{fig:NQ} (right), we plot as function of the jet 
hardness $Q$\footnote{The energy range $100\leq Q (\text{GeV})\leq200$
should be realistic for Tevatron and LHC phenomenology.}, 
the total average jet multiplicity (\ref{eq:total}), which accounts for
the primary state radiation off the heavy quark 
together with the decay products from the final-state flavoured hadrons, which
were introduced in section \ref{sec:defandnot}. 
For these predictions, we set ${\cal K}^{ch}=0.6$ in (\ref{eq:total}),
and $Q_0\sim\Lambda_{QCD}=230$ MeV \cite{Dokshitzer:1991wu}.
Moreover, the flavour decays 
constants $N_c^{dc}=2.60 \pm 0.15$ and $N_b^{dc}=5.55 \pm 0.09$ are independent
of the hard process inside the cascade, such that $N_A^{dc}$ can be added in the 
whole energy range. For instance, such values were obtained by the OPAL
collaboration at the $Z^0$ peak of the $e^+e^-$ annihilation.
In this experiment, $D^*$ mesons were properly reconstructed in order to provide
samples of events with varying $c$ and $b$ purity, such that it became possible
to measure light and heavy quark charged hadron multiplicities
separately \cite{Akers:1995ww}.
As compared to the average multiplicities of the primary 
state radiation displayed in Fig.\,\ref{fig:NQ} (left), after accounting for $N_A^{dc}$, the $b$ quark jet 
multiplicity becomes slightly higher than the $c$ quark jet multiplicity, although
both remain suppressed because of the dead cone effect. 

\section{Conclusions}

Thus, our present work focusing on the differences of the average charged
hadron multiplicity between jets initiated by gluons, light or heavy quarks
could indeed represent a helpful auxiliary criterion to tag 
such heavy flavours from background for jet hardness $Q \gtrsim 40$ GeV.
Notice that we are suggesting as a potential signature
the {\em a posteriori} comparison
between average jet multiplicities corresponding to 
different samples of events where other criteria
to discriminate heavy from light quark initiated jets
were first applied. 
Fig.\ref{fig:NQ} (right)  plainly demonstrate that the separation between light quark
jets and heavy quark jets is allowed above a few tens
of GeV with the foreseen errors of the experimentally measured average 
multiplicities of jets. The difference between light quark jet multiplicities 
and heavy quark jet multiplicities $N_q-N_Q$ in one jet is exponentially increasing because
of suppression of forward gluons in the angular region around the heavy quark direction. This
result is not drastically affected after accounting for heavy flavour decays multiplicities,
such that it can still be used as an important signature for $b$-tagging and the search 
of new physics in a jet together
with other selection criteria. 
Notice that the measurement of such observables require the previous reconstruction
of jets at hadron colliders \cite{Cacciari:2008gn}. 

\section*{Acknowledgements}

I gratefully acknowledge support from Generalitat Valenciana under grant PROMETEO/2008/004.
I also thank the organizers of Light Cone 2010, J. Papavassiliou and V. Vento, who gave
me the opportunity to present this work at the plenary session of this conference and finally,
my collaborators V. Mathieu and M.A Sanchis-Lozano.

\bibliographystyle{plain}

\bibliography{mybib}

\end{document}